**Characteristics of Molecular-biological Systems and Process-network Synthesis**

L. Papp*, S. Bumble**, F. Friedler* and L. T. Fan***

*Dept. of Computer Science, University of Veszprém, Hungary
**Dept. of Physics/Engineering, Philadelphia Community College, Philadelphia, PA 19130
***Dept. of Chemical Engineering, Kansas State University, Manhattan, KS 66506

**Abstract**

Graph Theoretic Process Network Synthesis is described as an introduction to biological networks. Genetic, protein and metabolic systems are considered. The theoretical work of Kauffman is discussed and amplified by critical property excursions. The scaling apparent in biological systems is shown. Applications to evolution and reverse engineering are construed. The use of several programs, such as the **Synprops**, **Design of molecules**, **Therm** and **Knapsack** are suggested as instruments to study biological process network synthesis. The properties of robust self-assembly and Self-Organizing synthesis are important contributors to the discussion. The bar code of life and intelligent design is reviewed. The need for better data in biological systems is emphasized.

Introduction
A. Graph Theory
B. Combinatorics
C. Process Network Synthesis
D. Biological Networks
　1. Exploration of Biological Networks through Analogy with Process Networks
　2. Genetic Networks
　3. Protein Networks
　4. Metabolic Networks
　5. Kauffman
　6. Critical Properties
　6a. Complexity
　7. Scaling
　7a. Small World
　8. Evolution
　9. Perturbation of Genetic and Protein Critical Networks
　9a. Correction of Protein Network Models With Protein Folding Phenomena
　10. Network Reconstruction or Reverse Engineering
E. Structure and Physical Properties
　1. Synprops: A Program for Relating Molecular Structure with Their Design
　2. Physical Properties from Groups
　3. Design of Molecules with Desired Properties by Combinatorial Analysis
　4. Therm
　5. Knapsack
　6. Minimization of Free Energy
　7. Reaction Pathway Network
　8. Robust Self-Assembly

9. Self-Organizing Systems
    F. The Bar Code of Life and Intelligent Design
    G. The need for Better Data
    H. Bibliography

**Introduction**

The study of networks pervades all of science, from neurobiology to statistical physics[1-6]. As Strogatz states in his review article[3], "Empirical studies have shed light on the topology of food webs, electrical power grids, cellular and metabolic networks, the world-wide web, the internet backbone, the neural network of the nematode worm Caenorhabditis elegans, telephone call graphs, coauthorship and citation networks of scientists, and the quintessential 'old-boy' network, the overlapping boards of directors of the largest companies in the United States. These databases are now easily accessible, courtesy of the internet. Moreover the availability of powerful computers has made it feasible to probe their structure; until recently, computations involving million-node networks would have been impossible without specialized facilities. Why is network anatomy so important to characterize? Because structure always affects function!"

The emerging network is more like a cell than a system of interlocking gears. The diverse components, each performing a specialized job, compose a dynamic system.

**A. Graph Theory[1-2]**

A graph is a finite set of nodes and arcs, where relations between nodes are represented by arcs. There are many different types of graphs. In Process Network Synthesis (PNS), described in C below, we use a unique bipartite graph called P-graph. It can handle two different types of nodes, commonly the set of operating units and the set of materials if it is used in Process Network Synthesis. Operating units are represented by horizontal bars and materials by circles. Arcs can exist between different types of nodes only. Direction of an arc represents the consumption or the production of a material by an operating unit, so that if the material is consumed by the operating unit, then the arc is directed to the operating unit, and if the material is produced by the operating unit, the arc is directed to the material.

**B. Combinatorics**

Combinatorics is a branch of mathematics. Its function is to generate certain groups of given elements and to determine the possible number of such groups. The mathematical programming approach to Process Network Synthesis has two important steps: the generation of the mathematical model and the solution of the model. Both steps have combinatorial aspects. In the first step graph representation of the mathematical model should be postulated, while in the second step the model to be solved contains combinatorial variables. Combinatorial algorithms have been developed to be able to solve even large scale industrial problems. Process Network Synthesis (PNS) is described below.

## C. Process-Network Synthesis,

## Combinatorial Problems in Chemical Engineering[7-10]

A number of papers by Friedler, Fan and their associates concern a major task in chemical engineering: the design of industrial plants. This is usually done by using a series of flowsheets to plan the best possible connection of the process units in specific ways that include the raw materials and products.

Up until just recently, we have never been certain that the best flowsheet or plan for the industrial plant or its revision, with constraints such as least cost, least maintenance, most profit, minimum pollution, etc., has been obtained. Now, however, a method leading to such a desired result is existent.

The task has been to develop a problem formulation that manifests the unique structure of the class of problems and a solution procedure that exploits the specific structure of the problem. The method requires an in-depth understanding of both the engineering and mathematical aspects of the problem. The algorithmic process synthesis requires as given: a set of products, a set of raw materials, and mathematical models of the operating units. Then, the best processes or every feasible process with optimality criteria of cost, waste generation, control, risk, and combinations of these factors can be generated. The rigorous technique is based on combinatorics.

Structural representation uses process graphs or P-graphs for unit operations in process synthesis. The axioms are: (S1) Every final product is represented in the structure; (S2) A material represented in the structure is a raw material if and only if it is not an output of any operating unit represented in the structure; (S3) Every operating unit represented in the structure is defined in the synthesis problem; (S4) Every operating unit defined in the structure has at least one path leading to a product; and (S5) If a material belongs to the structure, it must be an input to or an output from at least one operating unit represented in the structure.

The union of all combinatorially feasible structures is called the maximal structure. The generation of this maximal structure requires, for its synthesis, a set of raw materials, a set of products and a set of operating units. The output is the maximal structure which is the algorithmic generation of all solution-structures of the synthesis problem. The method utilizes the accelerated Branch and Bound (ABB) algorithm for solving PNS problems. It reduces the size of an individual subproblem through exclusion of those operating units that should not be included in any feasible solution of the problem. The method of generation of the optimal solution, by minimizing the number of subproblems to be solved, takes place with very great speed.

For 35 operating units, there are 34 billion subproblems. The five axioms reduce the 34 billion combinations of the operating units to 3,465 combinatorially feasible structures. The optimal solution is included in the set of 3,465 feasible structures. The P-graph concept is also used in separation problems, which include distillation columns, mixers, etc. The solution of all these problems can now often be done on a desk computer or PC in a couple of minutes.

A major goal of this work is to point out that some of the similarities between process network synthesis as applied to chemical engineering can be applied to biological networks and to suggest that some of the former techniques can lead to significant results in the study of biological networks. Also, some of the results of the physics of network theory and its applications to biological networks can be applied to chemical engineering flow sheets.

**D. Biological Networks**

Life is expressed as the interactions among different proteins and molecules. These form the cell but are not considered to be alive in themselves. Genetic networks modeled from the genes may include RNA, proteins and regulatory DNA sequences of nucleotides composing each gene. Such primary sequences are now known at an increasing rate. It is difficult to link the genes and their products with functional pathways and networks. Regulatory networks are functions of combinatorial matters that are very complex but details are emerging. The inputs and outputs of each node are becoming known, although their elucidation may be complex.

**D1. Exploration of Biological Networks through Analogy with Process Networks**

It is highly likely that the graph-theoretic method of synthesizing process networks outlined earlier can be adopted for exploring the generation and functioning of biological networks in the light of various analogies between these two classes of networks. Such analogies are self-evident: Every biological species comprises molecules and thus is a chemical species in forming a biological network. It is worth recalling that the graph-theoretic method of PNS (process-network synthesis) is based on P-graphs consisting of two types of vertices (nodes); one devoted by circles is for chemical species, and the other denoted by horizontal bars is for functional units, termed operating units, performing transformations. The arcs linking these two types of vertices signify the direction of the transformations. To aid the identification of analogies, we will describe programs later, to describe, document and solve for results and extents of pathways among species. This will break up the networks into steps that can be synthesized later to produce the network. We will also use a theoretical construction that aids in the analysis of steps or links in the networks that indicate which are the fundamental ones in the network or global process. Cellular functions can also be described as "modules" made up of many species of interacting molecules. Functional modules are biological pathways: collections of molecules, both large and small, that cooperate to perform a given function, such as protein synthesis, signal transduction, or the biosynthesis of small molecules.

**D2. Genetic Networks**

The dynamics of genetic networks rest upon time dependent connections. "Random Boolean networks" and global dynamics, which are organized into basins of attraction associated with them. Stability and adaptability that respond to perturbation are balanced and form an essential part of the design. Computer simulations and statistical mechanics are now modeling genetic structure. There is a robust nature to these systems and their genetic properties. The genetic regulatory system is a very large directed network. The components are the vertices and the edges proceed from the regulating to the regulated component

### D3. Protein Networks[11-16]

**"Lethality and centrality in protein networks[17-19]**

An article by Jeong, et al.[18], states, "Cell biology traditionally identifies proteins based on their individual actions as catalysts, signaling molecules, or building blocks of cells and microorganisms. Currently, we witness the emergence of a post-genomic view that expands the protein' role, regarding it as an element in a network of protein-protein interactions as well, with a 'contextual' or 'cellular' function within functional modules. Here we provide quantitative support for this paradigm shift by demonstrating that the phenotypic consequence of a single gene deletion in the yeast, S. cerevisiae, is affected, to a high degree, by the topologic position of its protein product in the complex, hierarchical web of molecular interactions.

The S. cerevisiae protein-protein interaction network we investigate has 1,870 proteins as nodes, connected by 2,240 identified direct physical interactions, and is derived from combined, non-overlapping data obtained mostly by systematic two-hybrid analyses. Due to its size, a complete map of the network, while informative in itself offers little insight into its large-scale characteristics. Thus, our first goal was to identify the architecture of this network, determining if it is best described by an inherently uniform exponential topology with proteins on average possessing the same number of links, or by a highly heterogeneous scale-free topology with proteins having widely different connectivities. The probability that a given yeast protein interacts with k other yeast proteins follows a power-law form. (This topology is also shared by the protein-protein interaction network of the bacterium, H. pylori). It has an approximate value of the degree distribution of ~2.5, with an exponential cutoff at $k_c \cong 20$. This indicates that the network of protein interactions in two separate organisms forms a highly inhomogeneous scale- free network in which a few highly connected proteins play a central role in mediating interactions among numerous, less connected proteins."

An important known consequence of the inhomogeneous structure in the network ís simultaneous tolerance against random errors coupled with fragility against the removal of the most connected nodes. Indeed, we find that random mutations in the genome of S. cerevisiae, modeled by the removal of randomly selected yeast proteins, do not affect the overall topology of the network. In contrast, when the most connected proteins are computationally eliminated, the network diameter increases rapidly.

The simultaneous emergence of an inhomogeneous structure in both metabolic and protein interaction networks indicates the evolutionary selection of a common large scale structure of biological networks, and strongly suggests that future systematic protein-protein interaction studies in other organisms will uncover an essentially identical protein network topology. The correlation between the connectivity and indispensability of a given protein confirms that despite the importance of individual biochemical function and genetic redundancy, the robustness against mutations in yeast is also derived from the organization of interactions and topologic position of individual proteins. Thus, a better understanding of cell dynamics and robustness will be obtained from integrated approaches that simultaneously incorporate the individual and contextual properties of all constituents of complex cellular networks.

Many drugs target single abnormal proteins. But proteins form networks, almost like a circuit board, that directs different cells. Those patterns can fluctuate minute-to-minute, depending on what medications or other compounds affect the proteins. And with cancer, a drug targeting just one protein may help some patients--but other times tumors just reroute themselves along the protein network to evade treatment. Because everything is hooked up in circuits, it gives cancer cells the advantage, a fantastic advantage.

## D4. Metabolic networks[22-24]

Biochemical reaction networks and their information pathways are known for 43 organisms. They are similar, robust, scale-free and error tolerant. Perhaps this precludes an inherent cellular life pattern. Metabolic reaction networks are rich topological designs. The vertices are designated as either substrates or molecular compounds. Their edges are designated as the metabolic reactions. They can be products of incoming edges (reactions) or educts of outgoing edges (reactions). The symbols $k_i$ or $k_0$ (number of incoming edges or outgoing edges) have an average path length of 3 and values of about 2.5 to 4. The degree is scale-free and obeys a power law with exponent about 2.2. The clustering coefficient is defined as the probability that the two nearest neighbors of a vertex are also nearest neighbors of each other. It is about 0.3 (much larger than the random coefficient) for Escherichia Coli and this organism has 282 nodes. Thus the neighborhood of the network vertex is a "clique". A graph-theoretical identification of pathways for biochemical reactions is referred to in section E7.

## D5. Kauffman[23]

Erdos and Renyi, 1959, showed, "as the ratio of edge to nodes, $E/N$ increases past 0.5, the size of the largest component jumps from very small to a giant component containing most of the nodes in the system. This is a first order phase transition. Intuitively, when the ratio of edges to nodes is small, only isolated nodes and small dyads or trees will exist. As more edges are randomly added, the small components will slowly grow larger. But when these become fairly large, a few more randomly added edges will connect these modest components into the giant component. Indeed, the phase transition occurs just when the ratio of the ends of edges equals the number of nodes".

Waldrop relates in his book[23a] about Kauffman's work, "any cell in an organism contains a number of regulatory genes that act as switches and can turn one another on and off. If genes can turn one another on and off, then you have genetic circuits." He further surmised that the genome as a whole could be simulated in such circuits and that the circuits could settle down into a stable, self-consistent pattern of active genes.

Now, the precise genetic details of any given organism would be a product of random mutations and natural selection, but the organization of life itself, the order, would be deeper and more fundamental. It would arise purely from the structure of the network, not the details.

As Kauffman states, "The genetic circuits you find in real cells have been refined by millions of years of evolution." Kauffman "shuffled the deck" and dealt out a bunch of utterly typical genetic circuits to see if they did produce stable configurations.

Regulatory genes were essentially just switches which flip back and forth between two states: active or inactive, similar to light bulbs (on-off) or as a statement in logic (true-false). These genetic light bulbs organize themselves into interesting patterns. The connections were to be sparse but not too sparse. Two-input networks always seemed to stabilize very quickly. He simulated a network with 100 genes. The network quickly settled into orderly states, with most of the genes frozen on or off and the rest cycling through a handful of configurations.

Kauffman "could tell that a system is at the critical state and/or the edge of chaos if it shows waves of change and upheaval on all scales and if the size of the changes follows a power law. One can expect any closely interacting population of agents to get themselves into a state of self-organized criticality, with avalanches of change that follows a power law." He performed this with a computer program that simulated a two-input network with 100 genes in it. This network quickly settled down into orderly states, with most of the genes frozen on or off and the rest cycling through a handful of configurations.

Kauffman's work seems to accentuate that the edge of chaos in such a system is very important in biological processes and that much activity proceeds from this state. Below is another description of what is called critical states, borrowed from physics, that also demonstrates how such processes can take place.

**D6. Critical Properties[24]**

A method for deriving the chemical potential of particles adsorbed on a two dimensional surface has previously been derived for lateral and next nearest neighbor interactions of the particles. In order to do so a parameter called $K$ was used and $K=((\exp((\mu-\varepsilon)/kT))\ (\theta/1-\theta))^{1/Z}$. It was found from a series of normalizing, consistency and equilibrium relations shown in papers by Hijmans and DeBoer (1) and used by Bumble and Honig (2) in a paper on the adsorption of a gas on a solid. In the above, $\mu$ is the chemical potential and $\varepsilon$ is the adsorption energy. The numerical values of $K$ were derived from computers for various lattices with different values of the interaction parameters for nearest neighbors ($c$), where $c=\exp(-w/kT)$, $w$ is the interaction energy, $Z$ is the coordination number with occupancy ($\theta$), the lattice occupation stands for the fraction of sites covered.

The value of $K$ given above has the chemical potential in it and solving for this quantity we obtain $K^Z$ ($\theta/1 - \theta$) where $Z$ is the coordination number of cells in a tissue approximated as a lattice.

The lattice has been approximated as triangular and was broken up into basic figures mentioned above. The larger the basic figure the more complicated the algebra. The bond yields $K=(\beta-1+2\theta)/(2\theta c)$, where $\beta=[1+4c\theta(1-\theta)(c-1)]$ and the triangle as basic figure yields a quartic equation $K^4 - a_1 K^3 + a_2 K^2 - a_3 K + a_4 = 0$ where $a_1=((2-5\theta)c+2-3\theta)/c(1-2)$, $a_2 = (c+5)/c$, $a_3 = ((3-5\theta)c+1-3\theta)/c(1-2\theta)$ and $a_4=1/c^2$. Every point then derived for the triangle basic figure subfigure is then found for the solution of the above quartic equation for given values of $c$ and $\theta$.

In previous applications of the above to gas adsorption, the value of $Z$ was fixed. However, here, in general network applications, $Z$ will vary. Thus, we vary $Z$ and find the value of $c$ ($c=\exp(-w/RT)$) when the isotherm just begins to be flat or parallel to the abscissa theta ($\theta$). When we do this we find that for the critical isotherms:

| $c$ | $Z$ |
|---|---|
| 2.3 | 4 |
| 1.72 | 6 |
| 1.33 | 12 |
| 1.22 | 18 |
| 1.17 | 24 |
| 1.13 | 30 |
| 1.11 | 36 |

Now when we plot $c$ versus $Z$ on log-log graph paper we obtain a straight line. This means the graph has scaling, which we will discuss later. Now $Z=4$ is a two dimensional square lattice, $Z=6$ is a two dimensional triangular lattice and $Z=12$ to 36 corresponds to a three dimensional sphere in which the central particle is in a plane surrounded by up to 36 particles in six slices or planes of six particles each in a hexagonal pattern; all particles of each plane are connected to the central particle (or interacts with the central particle). This occurs in three dimensions and furthermore the shell of surrounding particles in each plane are connected to each other, so that we can apply the triangular approximation discussed above, as well as the bond approximation to obtain the solutions.

This critical state corresponds to the state where the chemical potential remains constant. In this state (the grand canonical ensemble) particles will neither leave nor enter but reorder themselves or, in other terms, enter into the edge of chaos, which is a regime between order and disorder.

Sanchez, et al., state, "An investigation of the effect of directed links on the behavior of a simple spin-like model evolving on a small-world network was made. It showed that directed networks may lead to a highly nontrivial phase diagram including first and second-order phase transitions out of equilibrium."[25]

Newman, et al., relate, "A study of the theory of random graphs is shown with arbitrary distributions of vertex degree, including directed and bipartite graphs. Using the mathematics of generating functions, one can calculate exactly many of the statistical properties of such graphs in the limit of large numbers of vertices. Explicit formulas for the position of the phase transition at which a giant component forms, the size of the giant component, the average and distribution of the other components, the average number of vertices a certain distance from a given vertex, the clustering coefficient and the typical vertex-vertex distance on a graph all are shown."[26]

**6a. Complexity.** The Fall, 1999, Santa Fe Bulletin[27] has the following statements, "studying intermediate kinds of networks between lattice-like networks and random networks led Watts and Strogatz[27-28] to the discovery that when just a few long-range, random connections replace the local edges of a lattice-like network, the characteristic path length decreases dramatically; and a "shortcut" occurs. And while the first random rewiring has a great impact on path length, the clustering changes very little. Even when the separation of elements in a network is very small the clustering can remain almost as high as possible. The result is what Watts and Strogatz call a "small-world network." The name derives from the fact that it exhibits the short global separations that are typified (anecdotally) by social interactions while maintaining the high degree of clustering exhibited in most social networks. The neural network of the nematode worm C. elegans is the sole example of a completely mapped neural network and turns out to be a small-world network."

**D7. Scaling**

Many large networks are scale-free (SF); their degree distribution follows a power-law for large k, i.e., (nodes that have many connections). SF networks come about because they model network assembly and evolution. When we capture correctly the processes that assembled the networks, then we will obtain the topology as well. Most real networks exhibit preferential attachment. Growth and preferential attachment caused the introduction of the SF model that has a power-law degree distribution.

**7a. Small World**. As big as it is, this is a "small world". The "small world" of social networks is known as "six degrees of separation". Small clusters of social networks each node connected to the world outside their own circle of friends with a few weak links. The 800 million documents (or nodes) of the web (up to 1999) has a shortest path of only 19 links.

According to Kim, et al.[29], "the possibility of the small-world phenomenon in scale-free networks subjected to several local strategies of path finding was found. A very simple local strategy based on local connectivity information leads to the scaling behavior $D \sim \log_{10} N$. The error tolerance and the attack vulnerability found previously by global strategy have been shown to be generic topological properties of scale-free networks, and do not depend on the path finding strategy."

**D8. Evolution**

**The Time Machine**

The Graph Theoretic Process Synthesis Program, described in part C above, functions as a time machine. It propels the very intelligent engineer towards the far distant best design which he could have arrived at, had he practically infinite time. Thus it is in nature. A systematic comparative mathematical analysis of the metabolic and protein networks of 43 organisms representing all domains of life have the same organization of robust and error-tolerant scale-free networks representing a common blueprint for large-scale organizations of all cellular constituents.

Originally it was thought that complex networks were modeled using the random network theory introduced by Erdos and Renyi, where each pair of nodes in the network is randomly connected, leading to a network where most nodes have the same number of links. Thus the probability of finding a highly connected node decays exponentially (p($k$) is equivalent to $e^{-k}$) for $k>>\{k\}$. However it has been found that the World-Wide Web, the internet and social network systems are described by a few highly connected nodes which link the rest of the less connected nodes to the system (p($k$) is equivalent to $k^{-\gamma}$). Now it has been found that this is also true for the biological networks referred to above.

This must be so because they adhere to Darwin's theory of evolution. After many millenniums the organisms have been adapting to their environment and the survival of the fittest has led to the structure of their metabolic and protein networks that are most fit and are most robust and error-tolerant scale-free networks.

The graph theoretic process synthesis flowsheet results may be of that nature, i.e., if one tried to fit the log p($k$) versus the log $k$, the curve would be linear and it would not follow a Poisson distribution that peaks sharply at $\{k\}$ and decays exponentially from the value of $\{k\}$.

It is to be observed that the word "optimum" was not used. In the biological sense the systems have not found the optimum. They have only found the best fit to the environment. In the process synthesis sense, likewise, the optimum is not really found, only the best fit to the constraints, be they cost minimized or profit maximized. In the nature case, for the adaptation of the network system to the environment, the method is not the most efficient, but proceeds by trial and error, and after a suitable time period, to find the best fit. Therefore in genetic and protein networks, we must attempt to simulate, rather than optimize, the networks of organisms with computer programs.

An article by Podani, et al.[30], relates, "the comparison of information transfer pathways demonstrates a close relationship of Archaea and Eukarya. Similarly, although eukaryotic metabolic enzymes have been shown being primarily of bacterial origin, the pathway-level organization of archaeal and eukaryotic metabolic networks also proves more closely related. Our analyses therefore reveal a similar systemic organization of Archaea and Eukarya, and suggest that during symbiotic origin of eukaryotes the incorporation of bacterial metabolic enzymes into the proto-archaeal proteome was constrained by the host's preexistent metabolic architecture. The large-scale metabolic organization is essentially identical in all contemporary species, all possessing a robust and error tolerant scale-free network architecture. The complete metabolic network is under organizational constraints. A selective pressure during the symbiotic evolution of eukaryotes must have been present that limited their incorporation of bacterial metabolic enzymes, in order to maintain the already existing near optimal metabolic network architecture inherited from the proto-archaean host."[30]

Ipsen, et al.[31], state, "an evolutionary algorithm for reconstruction of graphs from their Laplacian spectra is presented Through a stochastic process of mutations and selection, evolving test networks converge to a reference graph. The method allows exact reconstruction of relatively small networks and yields good approximations in the case of large sizes for random and clustered graphs and small-world networks. This is a learning process where a test network, by adjusting its internal structure, learns to approximate dynamics generated by a different system. Approximations of large clustered graphs by graphs of smaller size can be constructed and networks that generate stochastic signals with power spectra can be designed."[31]

Cancho, et al.[32], state, "when dealing with biological networks, the interplay between emergent properties derived from network growth and selection pressures has to be taken into account. As an example, metabolic networks seem to result from an evolutionary history in which both preferential attachment and optimization are present. The topology displayed by metabolic networks is scale-free, and the evolutionary history of these nets suggests that preferential attachment might have been involved. Early in the evolution of life, metabolic nets grew by adding new metabolites, and the most connected are known to be the oldest ones. On the other hand, several studies have revealed that metabolic pathways have been optimized through evolution in a number of ways. This suggests that the resulting networks are the outcome of both contributions, plus some additional constraints imposed by the available components to the evolving network. In this sense, selective pressures might work by tuning underlying rules of net construction. This view corresponds to Kauffman's suggestion that evolution would operate by taking advantage of some robust, generic mechanisms of structure formation."[32]

**D9. Perturbation of Genetic and Protein Critical Networks**

A method is suggested to design new drugs and counteract disease that builds upon the combinatorial method now existent at drug companies, yet incorporates the matter introduced in this paper on protein networks. To introduce this method, let us first review the combinatorial method now existent.

**Method Now Used[33]**

We wish to inhibit an enzyme (BAD) that causes certain cells to physically react, so that a bad health effect occurs in the body. We will call a drug that inhibits this action BAD-1. A search is then conducted in the huge public genome database for snippets of DNA that resemble the known BAD enzyme. From this search, 10,000 computer matches are obtained, each of them gene fragments representing potential BAD inhibitors. Let us suppose this snippet is TGCAATG. Then the library of several million frozen DNA fragments is explored and those that correspond to these computer matches are placed in tiny wells on plastic plates and fed to a robot machine called Zeus. Zeus dips tiny needles into the wells, picks up microscopic droplets of DNA and drops them onto a sheet of nylon paper, creating a so-called microarray. These sheets are rolled up and slipped into test tubes. The test tubes are then washed with genetic material from a wide range of tissue cells that have been labeled with a radioactive dye. When a gene is active in a particular cell type, the spot lights up under UV. By comparing patterns of brightness, the researchers isolate one gene, which they call BAD-2 that looks promising. Confirming that BAD- 2 does indeed convert the enzyme, the researchers look for a chemical that will inhibit BAD-2. They have a head start because they know what BAD inhibitors look like. In their library of molecules they find one that works against BAD-2. This is their drug candidate. However, although the drug companies formerly thought this method was their *sine qua non* for finding drug candidates, they now are looking for better methods to find lead compound drugs.

**Extension of the Method Now Used**

It is proposed that an extension to the method now used be extended by changing the word DNA referred to above to the word protein. The snippet will now be changed to, e.g., a sequence of amino acids that is a fragment of the protein that is sought. This could be AGPWLVM. This method is then analogous to the method above except that we are now dealing with Proteomics instead of DNA. There are many more proteins and sequences of amino acids possible than there are genes or sequences of nucleotides, so that much work would remain to be done in building up the library necessary for this method, but such work is proceeding.

**A New Method**

A new method is herein proposed that utilizes critical network theory described above. Since complex systems are constructed so that they are on the boundary between order and chaos, these are the systems best able to adapt by mutation and selection. In this method a candidate or LEAD NETWORK is found from the network that best satisfies the following constraints:

The vertices (or proteins) each have a combined coordination number and interaction energy that best satisfy the conditions for criticality established in D6, have physical properties that correspond with results (as indicated by Synprops with the best data), compare closely with the distribution of amino acids from experimental results (as indicated also by Knapsack and Synprops) and agree with QSARs (quantitative structure activity relationships). Furthermore, the adjacency matrix is composed of elements that will meet the constraints above as closely as possible, the local aspects of the network can be extended to the global aspect and finally matrix equations combining all of these constraints can be solved for their component elements as was done as a byproduct result of the Synprops method.[35]

The LEAD NETWORKS will then not only be instrumental in finding the root of the disease but also in fighting the disease itself.

**9a. Correction of Protein Network Models with Protein Folding Phenomena.** Kramers published his "pearl necklace model" of polymer molecules over 50 years ago (J. Chem. Phys., 14, 415, 1946). This can be used to find the angles between the links of a polymer, each link of a different mass and as a function of kinetic and potential energy, using statistical physics. Since it has been shown that once the configuration of three particular links in the protein chain are ascertained, the whole polymer configuration can be found, then the corrections for the approximation of considering the protein as a point rather that the valid folded functional polymer can be made. It is also possible to do so because the polymer configuration can be put into its particular set of evolutionary forms that the protein will assume. Then the vertices of the network can be more realistically portrayed and the interactions between them can be arrived at with greater precision and accuracy.

**D10. Network Reconstruction and Reverse Engineering**

"**Models of Genetic Networks**

In the preceding three decades, theoretical models of genetic regulatory networks have been studied intensively. Since only sparse data on eukaryotic gene regulation were available, most studies were based on highly idealized models such as networks with random links connecting the nodes and randomly chosen Boolean rules. With the arrival of high-throughput technologies the focus has recently shifted towards more realistic models. The reconstruction of genetic networks from time resolved measurements of these gene expression matrices belongs to the class of inverse problems and is given the name of reverse engineering.

Each of the vertices of biological networks, whether it is a protein network or a metabolic network, may represent a chemical species and thus has physical properties. Methods that convert structure to properties are described below. Methods are also described that convert properties and hence the structure or vertices to an optimum or constrained values."

**E1. Synprops[34]**

Cramer's data was portrayed in a table of group properties. Results so obtained were from extensive regressions on experimental data from handbooks and were tested and statistically analyzed. The data was used to predict physical properties for other compounds than those used to derive the data. Optimization procedures were combined with the Cramer data (in an extended spreadsheet), and applied for Pollution Prevention and Process Optimization. In addition, Risk Based Concentration Tables, etc., were included as constraints to insure that the resulting composite structures were environmentally benign.

During the course of many years, scientists have recognized the relationship between chemical structure and activity. Pioneering work had been done by Hammett in the 1930s, Taft in the 1950s and Hansch in the 1960s. Brown also recognized the relation between steric effects and both properties and reactions. QSAR methodologies were developed and used in the areas of drug, pesticide and herbicide research. In the 1970s, spurred by the increasing number of chemicals being released to the environment, QSAR methods began to multiply.

**E2. Physical Properties form Groups**

It has also been known that a wide range of properties can be derived using The Principle of Corresponding States which used polynomial equations expressed with reduced temperature and pressure. In order to obtain the critical properties needed for the reduced temperature and reduced pressure, the critical constants are derived from the parameters for the groups of which the molecules are composed.

Thus, the treatment of many molecules through their composite groups and the connection with their properties becomes an exercise of obtaining good data to work with. This is particularly difficult for drug and ecological properties that are not in the public domain.

Cramer's method consisted of applying regressions to data from handbooks, such as the Handbook of Chemistry and Physics, etc., to fit the physical properties of molecules with the groups comprising their structures. The results considered about 35 groups and were used in the Linear-Constitutive Model and a similar number of groups (but of a different nature) were used in the Hierarchical Additive-Constitutive Model. Statistically a good fit was found and the prediction capabilities for new compounds were found to be excellent.

Twenty-one physical properties were fitted to the structures. The Properties (together with their dimensions) were Log activity coefficient and Log partition coefficient (both dimensionless), Molar refractivity ($cm^3$/mol), Boiling point (degrees C.), Molar volume ($cm^3$/mol), Heat of vaporization (kcal/mol), Magnetic susceptibility (cgs molar), Critical temperature (degrees C.), Van der Waals $A^{1/2}$ (L $atm^{1/2}$/mol), Van der Waals B (L/mol), Log dielectric constant (dimensionless), Solubility parameter (cal/$cm^3$), Critical pressure (atm), Surface tension (dynes/cm), Thermal Conductivity ($10^4$x($cals^{-1}cm^{-2}$(cal/cm)$^{-1}$), Log viscosity (dimensionless), Isothermal ($m^2$/molx$10^{10}$), Dipole moment (Debye units), Melting point (degrees C.) and Molecular weight (g/mol). Later the equations for molar volume (Bondi scheme) and molar refractivity (Vogel scheme) were included as were also equations for the Log Concentration X/Water, where X was ether, cyclohexane, chloroform, oils, benzene and ethyl alcohol, resp., Risk-Based Concentrations and Biological Activity equations were also included.

The formulas for physical properties, activities or Risk Based Concentrations is the general linear combination equation

$$P_{ij} = a_i + b_i B_j + c_i C_j + d_i D_j + e_i E_j + f_i F_j$$

The *i* subscripts stand for different properties and the *j* subscripts indicate different molecules. It may be seen that the spreadsheets are like the blueprints of a molecule whose structure is the composite of the numbers in one column and whose properties are given in another column. The quantities *B...F* are analogous to the genes (5 in this case) in living systems.

The hardware and software for personal computers have developed very rapidly. Thus the treatment of many molecules through their composite groups and the connection with their properties became an exercise of obtaining good data to work with. In the PC spread sheet "Tools" part of the program is an Optimizer program, which was used in this work. The technology of the modern PC was matched with the power of mathematics to obtain results. Thus the structure of molecules could be constrained by their physical properties or molecules could be designed with the best physical properties of concern.[35]

The physical properties of the links in the protein molecule were assessed from the Synprops program. The following eight properties were of particular interest: dipole moment, volume, hydrophobicity, log of the activity coefficient, log of the partition coefficient, solubility parameter, chi, and molar refractivity. In the literature, it is stated that the hydrophobicity is most important. Contrary to this it has been surmised through the comparison from results from the Synprops program, the Knapsack program and comparison of the literature values for the frequencies of amino acid residues of the genomes of living organisms, Archaea, Bacteria, Eukaryotes, etc., that the importance of the properties rank as follows: log of the partition coefficient, hydrophobicity, volume, dipole moment. It is also important to note that many times these physical properties are functions of each other. In fact the log of the partition coefficient can be derived very closely as the function of two other properties. A false route that can be taken is the tracing of the relation of the properties to the evolution of organisms. This is because many organisms are not developed from others but can be branches of the same evolutionary hierarchical routes.

**E3. Design of Molecules with Desired Properties by Combinatorial Analysis[36]**

Friedler and Fan and their associates further published work that stated: "Suppose that a set of groups to be considered and the intervals of values of the desired properties of the molecule to be designed are given. Then, the desired properties constitute constraints on the integer variables assigned to the groups. The feasible region defined by these constraints is determined by an algorithm involving a branching strategy. The algorithm generates those collections of the groups that can constitute structurally feasible molecules satisfying the constraints on the given properties. The molecular structures can be generated for any collection of the functional groups.

The proposed combinatorial approach considers only the feasible partial problems and solutions in the procedure, thereby resulting in a substantial reduction in search space. Available methods exist in two classes. One composes structures exhaustively, randomly or heuristically, by resorting to expert systems, from a given set of groups; the resultant molecule is examined to determine if it is endowed with the specified properties. This "generate-and-test" strategy is usually capable of taking into account only a small subset of feasible molecular structures of the compound of interest. It yields promising results in some applications, but the chance of reaching the target structure by this strategy can be small for any complex problem, e.g., that involving a large number of groups. In the second class, a mathematical programming method is applied to a problem in which the objective function expresses the "distance" to the target. The results of this assessment may be precarious since the method for estimating the properties of the structure generated, e.g., group contributions, is not sufficiently precise. The work here is a combinatorial approach for generating all feasible molecular structures, determined by group contributions, in the given intervals. The final selection of the best structure or structures is to be performed by further analysis of these candidate structures with available techniques."

**E4. THERM[37]**

A set of subprograms collectively called THERM written by Ritter and Bozelli[37] have proven very useful in programs for equilibrium such as the minimum free energy program and those for chemical kinetics. THERM supplies the thermochemical data needed by these two programs based on the data files in Benson's book Thermochemical Kinetics.[38] The data in THERM is given for 330 groups. They are divided into 3 types of groups, BD-bond dissociation groups, CDOT-radical groups and Regular groups-all other groups with no unbonded electrons. The last groups consist of HC-hydrocarbon groups, CYCH-ring corrections for hydrocarbon, oxygen, and nitrogen containing ring systems, INT-interaction groups/substituent effects, CLC-chlorine and halogen containing groups, HCO-oxygen containing groups and HCN-nitrogen containing groups. Recipes guide the user into the various subprograms. ENTER/ESTMATE SPECIES asks for species ID, number of groups in the species, number of different groups to be entered, elemental formula, groups contained in the species and the species symmetry number. (The code for the groups in the species takes a while to get used to). The result when stored in *Therm.lst* is in the form heat of formation, entropy and heat capacity at 300, 400, 500, 800, 1000, and 1500 degrees Kelvin. The whole *Therm.lst* file for many species can be converted to the *Therm.dat* file, which is exactly what is needed in the NASA format for LSENS and Chemkin programs. This is done with amazing speed by the subprogram *Thermfit*. Incidentally, literature values for H, S and Cp can be converted to the NASA format with *Thermfit* as well. The *Therm.dat* file is also needed in *Thermrxn* which is a sub program where a chemist writes a reaction which is immediately converted to a table of enthalpies entropies free energies, equilibrium constants, ratio of Arrhenius factors at a series of temperatures and in units of one's option. All the information found by THERM can be sorted and stored in specific files easily recalled.

A protein is described by the series of amino acid residues attached to its backbone. A method of determining the frequency of occurrence of the remnants of these amino acids in a given protein is described below. A combinatorial method such as that described in E5 may then be used to correlate such frequencies of the constituents of the proteins of various organisms to their evolution.

**E5. The Knapsack Problem**

It would be interesting to use physical properties to study the folding of proteins which leads to their function. Accordingly, it was thought that the properties of volume and hydrophobicity would be of use in that hydrophobicity is preferred in the folding and also the extended version of the polymer configuration changes to a globular packed version before its function can be realized as a target for docking. The question then arises as to how many ways can one pack a fixed container with property values without exceeding, but fully utilizing, the space that is preordained for the container.

It then occurred that the knapsack methodology of discrete mathematics could be used. The knapsack problem for solutions arises when a single critical resource is to be optimally allocated among a variety of options. Specifically, a number of items are available, each of which consumes a known resource and contributes a known benefit. Items are to be selected to maximize the total benefit without exceeding the given amount of the resource. If $N=\{1, 2,…,n\}$ is a given set of $n$ items and $j$ consumes (requires) $a_j>0$ units of the resource and confers the benefit $c_j>0$, we have the knapsack problem as the following 0-1 integer linear programming problem:

maximize: $\Sigma_{j \in N} c_j x_j$

subject to: $\Sigma_{j \in N} a_j x_j$ is smaller than or equal to $b$

$x \in \{0,1\}$

where $a_j$ is smaller or equal to $b$ for all $j \in N$.

Thus proteins could be designed for frequencies of amino acid residues that imbues the protein with the best desired physical properties or the structure that conforms most to the properties desired. Indeed, this can be carried out for all the properties, not only hydrophobicity or volume. The values of hydrophobicity and volume were obtained from Synprops (as were some 30 other properties) but only six other chosen properties were singled out for special attention: dipole moment, log partition coefficient, log activity coefficient, chi, molar refractivity and solubility parameter for fifteen amino acid residue links of the polypeptide. The eight were reduced to seven because the molar solubility parameter and the log activity coefficients were found to be functions of each other. Later the volumes and hydrophobicity were found in the literature[39-40] for all 20 amino acids and they were also used in the program. The values for the latter two properties of the amino acids did not agree for the values from Synprops with those from the literature for these two properties, but when plotted, the curves were isomorphic with each other except for the value for hydrophobicity of lys. Conversion for Knapsack usually is less than one second. Knapsack has been very useful for studying the evolution of simple organisms from their frequency of occurrence of amino acid residues. "Empirical programs" such as Chemkin exist to follow the kinetics of chemical reactions to their final results.

**E6. Minimization of Free Energy as formulated by White and Dantzig[41]**

White and Dantzig work states: When reliable and consistent thermodynamic data are available, equilibrium calculations of biochemical species become accessible. The method of the minimization of free energy was pioneered by the Rand Corporation for the most efficient method for this. In this method conservation of mass m gaseous elements used to form compounds under constant pressure $P_i$ with $a_{ij}$ atoms of the $i$th element in one molecule of the $j$th compound, and with $b_i$ moles of the $i$th element and $x_j$ moles of the $j$th compound in the mixture yields:

$$\sum_{j=1}^{n} a_{ij} x_j = b_i$$

with $i = 1, \ldots, m$

and $x_j$ equal to or greater than 0 and $n$ equal to or greater than $m$. Subject to these constraints it is desired to minimize the total Gibbs free energy of the system:

$$\sum_{j=1}^{n} c_j x_j + \sum_{j=1}^{n} x_j (\log(x_j / \sum_{i=1}^{n} x_i))$$

where $c_j = F_j/RT + \log P$
$F_j$ = Gibbs energy per mole of the $j$th gas at temperature $T$ and unit atmosphere pressure
$R$ = universal gas constant

## E7. Reaction Pathway Network[42]

Drs. Fan and Friedler and associates[42a] have a method for selecting candidate pathways or mechanisms of a reaction through synthesis of networks of elementary reactions. This offers a way to reduce the number of reactions that may occur in the Chemkin input and also a way to translate, more simply, the resulting reduced number of reactions into a network. This can yield a complete network, i.e., the maximal network, for a given set of plausible elementary reactions based on the combinatorial properties of the reaction pathway. The subnetworks containing intermediaries as products in addition to targets of the reaction of concern (as byproducts) are eliminated. The remaining networks that satisfy element conservation are finally selected as the feasible candidate networks or mechanisms. The number of these mechanisms is usually much smaller, greatly verifying the mechanism. The method is firmly rooted in a set of combinatorial axioms and expressed in the terminology of process graphs or P-Graphs. The conversion of glucose to pyruvate with 14 elementary reactions follows a method for synthesizing networks of metabolic pathways by this highly exacting combinatorial method[42b].

## E8. Robust Self-Assembly Using Highly Designable Structures and Self-Organizing Systems[35]

Through a statistical exploration of many possibilities, self-assembly creates structures. These explorations may give rise to some highly designable structures that can be formed in many different ways. If one uses such structures for self-assembly tasks, a general approach to improving their reliability will be realized.

Manufacturing builds objects from their components by placing them in prescribed arrangements. This technique requires knowledge of the precise structure needed to serve a desired function, the ability to create the components with the necessary tolerances and the ability to place each component in its proper location in the final structure.

If such requirements are not met, self-assembly offers another approach to building structures from components. This method involves a statistical exploration of many possible structures before settling into a possible one. The particular structure produced from given components is determined by biases in the exploration, given by component interactions. These may arise when the strength of the interactions depends on their relative locations in the structure. These interactions can reflect constraints on the desirability of a component being near its neighbors in the final structure. For each possible structure the intersections combine to give a measure of the extent to which the constraints are violated, which can be viewed as a cost of energy for that structure. Through the biased statistical exploration of structures, each set of components tends to assemble into that structure with the minimum energy for that set. Thus, self-assembly can be viewed as a process using a local specification, in terms of the components and their interactions, to produce a resulting global structure. The local specification is, in effect, a set of instructions that implicitly describes the resulting structure.

Self-assembly can form structures beyond the current capacity of direct manufacturing. The most straight forward technique for designing self-assembly is to examine with a computer simulation, the neighbors of each component in the desired global structure, and then choose the interactions between components to encourage these neighbors to be close together.

A difficulty in designing the self-assembly process is the indirect or emergent connection between the interactions and the properties of resulting global structures. There is a possibility of errors due to defective components or environmental noise. To address this problem, it would be useful to arrange the self-assembly so the desired structure can be formed in many ways, increasing the likelihood they will be correctly constructed even with some unexpected changes in the components or their interactions. That is, the resulting global structure should not be too sensitive to errors that may occur in the local specification.

A given assembly can then be characterized by the number of different component configurations producing a given global designability. Self-assembly processes with skewed distributions of designability can also produce relatively large energy gaps for the highly designable structures. A large energy gap with small changes in the energies of all the global structures do not change the one with the minimum energy, but small changes with a small gap are likely to change the minimum energy structure. If there are several structures that adjust reasonably well to the frustrated constraints in different ways, the energy differences among these local minima will determine the gap.

Self-assembly of highly designable structures is particularly robust, both with respect to errors in the specification of the components and environmental noise. Thus we have a general design principle for robust self-assembly: select the components, interactions and possible global structures so the types of structures desired for a particular application are highly designable.

Applying this principle requires two capabilities. The first is finding processes leading to highly designable structures of the desired forms. The second requirement is the ability to create the necessary interactions among the components.

Achieving a general understanding of the conditions that give rise to highly designable structures is largely a computational problem that can be addressed before actual implementations become possible. Thus developing this principle for self-assembly design is particularly appropriate in situations where explorations of design possibilities take place well ahead of the necessary technological capabilities. Even after the development of precise fabrication technologies, principles of robust self-assembly will remain useful for designing and programming structures that robustly adjust to changes in their environments or task requirements.

**E9. Self-Organizing Systems (SOS)[35]**

Some mechanisms and preconditions are needed for systems to self-organize. The system must be exchanging energy and/or mass with its environment. A system must be thermodynamically open because otherwise it would use up all the available usable energy in the system (and maximize its entropy) and reach thermodynamic equilibrium.

If a system is not at or near equilibrium, then it is dynamic. One of the most basic kinds of change for SOS is to import usable energy from its environment and export entropy back to it. Exporting entropy is another way to say that the system is not violating the second law of thermodynamics because it can be seen as a larger system-environment unit. This entropy-exporting dynamic is the fundamental feature of what chemists and physicists call dissipative structures. Dissipation is the defining feature of SOS.

The magic of self-organization lies in the connections, interactions and feedback loops between the parts of the system, it is clear that SOS must have a large number of parts. These parts are often called agents because they have the basic properties of information transfer, storage and processing.

The theory of emergence says the whole is greater than the sum of the parts, and the whole exhibits patterns and structures that arise spontaneously from the parts. Emergence indicates there is no code for a higher-level dynamic in the constituent, lower-level parts. Emergence also points to the multiscale interactions and effects in self-organized systems. The small-scale interactions produce large-scale structures, which then modify the activities at the small scales. For instance, specific chemicals and neurons in the immune system can create organism-wide bodily sensations which might then have a huge effect on the chemicals and neurons. Prigogine has argued that micro-scale emergent order is a way for a system to dissipate micro-scale entropy creation caused by energy flux, but this is still not theoretically supported.

Even knowing that self-organization can occur in systems with these qualities, it's not inevitable, and it's still not clear why it sometime does. In other words, no one yet knows the necessary and sufficient conditions for self-organization.

## F. BarCode and Intelligent Design[43]

Genetics originally discussed the nature of DNA, its constitution, organization and function and how a code leads to proteins. It is the nature and the function of the myriad of proteins to lead to the essence of life. The process proceeds through the formation of a set of chemical compounds called amino acids using a coding that is known as the genetic code. The amino acids are then arranged in a sequence that is permuted, seemingly haphazardly, but actually by intelligent design on a backbone of a peptide polymer or polypeptide and we call this structure a protein. A study of these proteins is known as proteomics. Proteins can change their form from stretched out string- like molecules to fold into globule forms and in these forms can bring about many processes that are necessary to life.

One protein is be able to "recognize" another protein. The characteristics of proteins are considered as properties and by that is mean the very physical, chemical and biological properties that are described in SYNPROPS and others that have not been considered in that program. It would seem that this recognizability is portrayed in the "shape space" of the proteins which is the description of a protein portrayed in a multi dimensional space, where the coordinates are properties rather than coordinates. Perhaps there is a morphic resonance or a morphogenetic sense set up by the proteins as in a "Sheldrake[44-45] sense" that is "perceived by each protein and that is recognized by another protein. This field, whether set up by signaling or in another manner, is perceived by a protein as specific to another protein. This, of course, is an alternate theory that contrasts with the "lock and key" concept or docking concept that may help to lead to new drugs for battling or curing disease.

Berkovich[43] has suggested, "the genome information plays the role of a "barcode". Thus the DNA structure is a pseudo-random number (PRN) with classification tags, so organisms are characterized by DNA as library books are characterized by catalogue numbers. Elaboration implicates the infrastructure of the physical Universe as a seat of biological information processing. The PRN allows the biological objects to share facilities in the Code Division Multiple Access (CDMA) mode, similarly to cellular phone connections. A community of users shares the phenomenon of Life as a collective information processing activity and this is treated as a methodology of engineering design. The concept of the "barcode" functionality of DNA confronts the descriptive scientific doctrines with a unique operational scheme of biological information control. The vital constituent of life - unfolds as extracorporeal information processing in the cellular automaton infrastructure underlying the physical world. These PRNs in nature serve the same purpose as SSNs in society - unique identification of systems' elements. A new organism starting from a zygote is equipped by its DNA with a unique identification number for opening "an account" on the "Internet of the Physical Universe".

## G. The Need for Better Data

In this review of work the need for better data is stressed. More data, as accurate as possible, is necessary to substantiate the importance of programs such as Synprops, Knapsack and the Design of Molecules, Whether this data comes from Therm, NIST, or other sources, such data is important to us for the work of achieving success in beneficial drug discovery and the understanding of life processes.

**Conclusions**

A number of programs exist that can characterize biological networks. The analogy between biological networks and Graph Theoretic Process Network Synthesis is displayed. A model of the critical regions of lattices is used to discuss systems between order and disorder of biological networks. A number of programs exist, such as Synprops, Therm, Knapsack, and those for the molecular design of molecules with specific properties that can elucidate biological networks. However, the physical and thermodynamic data that is accurate and consistent for biological molecules necessary for these programs is lacking and efforts must be made to fulfill the need for reliable data.

**H. Bibliography**

1. Brian Hayes, Graph Theory in Practice: Part I. American Scientist, Computing Science: January- February 2000
2. Brian Hayes, Graph Theory in Practice: Part II, American Scientist: Computing Science: March-April 2000
3. Steven H. Strogatz, Exploring Complex Networks, Insight Review Articles, Nature. Vol. 410, March 8, 2001
4. Reka Albert, Barabasi, Albert-Laszlo, Statistical Mechanics of Complex Networks, arXiv: cond-mat/0106096 v1, June 6, 2001
5. Dorogovtsev, S. N., Mendes, J. F. F., Evolution of Networks, arXiv: cond-mat/0106144
6. Physics World, The Physics of the Web, Volume 14, Issue 7, July 2001
7. Friedler, F., Tarjan, K., Huang, Y., W., Combinatorial Algorithms for Process Synthesis, Computers Chem. Engng., Vol. 16, Suppl., pp.S1-S548, 1992
8. Friedler, F., K. Tarjan, Y. W. Huang, L. T. Fan, Graph-Theoretic Approach to Process Synthesis: Axioms and Theorems, Chemical Engineering Science, Vol. 47, No. 8, pp. 1973-1988, 1992
9. Friedler, F., K. Tarjan, Y. W. Huang, L. T, Fan, Graph-Theoretic Approach to Process Synthesis: Polynomial Algorithm for Maximal Structure Generation, Computers Chem. Engng. Vol. 17, No. 9, pp. 929-942, 1993
10. Friedler, F., J. B. Varga, L. T. Fan, Decision mapping: A Tool for Consistent and Complete Decisions in Process Synthesis, Chemical Engineering Science, Vol. 50, No. 11. Pp. 1775-1768, 1995
11. Bernhard Palsson, The Challenges of in Silico Biology, 2000 Nature America Ing., http://biotech.nature.com
12. Hawoong Jeong, Barabasi, Albert-Laszlo, The Global of Cellular Networks
13. Bas Dutilh, Analysis of Data From Microarray Experiments, The State of the Art in Gene Network Reconstruction, Literature Thesis, Theoretical Biology and Bioinformatics, Utrecht University
14. Patrik D'haeseleer, Liang, S., Somogyi, R., Gene Expression Data Analysis and Modeling, Session on Gene Expression and Genetic networks, Pacific Symposium on Biocomputing, Hawaii, Jan. 4-9, 1999
15. Harley H. McAdams, Arkin, A., Simulation of Prokaryotic Genetic Circuits, Ann. Rev. Biomol. Struct. 1998. 27:199-224
16. A. Wuensche, Genomic Regulation Modeled as a Network with Basins of Attraction, http://www.santafe.edu/~wuensch/ , http://www.santafe.edu/~wuensch/ddlab.html
17. Dimitri Gilis, Massar, S., Cerf, N. J., Rooman, M., Optimality of the Genetic Code With Respect to Protein Stability and Amino-acid Frequencies, Genome Biology 2001, 2(11): research0049.1-0049.12, October, 2001
18. H. Jeong, Mason, S. P., Barabasi, A.-L., Oltvai, Z. N., Lethality and Centrality in Protein Networks, Nature, Vol. 411, May 3, 2001
19. Jeff Hasty Collins, J. J., Protein Interactions: Unspinning the Web, Nature, 411, 30-31 (2001)
20. Letter to Nature, H. Jeong, B. Tombor, R. Albert, Z. N. Oltvai & A.-L. Barabasi, The Large- Scale Organization of Metabolic Networks, Nature, Vol. 407, 5 October 2000, http://www.nature.com

21. Markus W. Covert, Schilling, C. H., Famili, I., Edwards, J. S., Goryanin, I. I., Selkov, E., Pallson, B. O., Metabolic Modeling of Microbial Strains in Silico, Trends in Biochemical Sciences, Vol. 26, No. 3, March 2001
22. Christopher H. Schilling, Palsson, B. O., The Underlying Pathway Structure of Biochemical Reaction Networks, Proc. Natl. Acad. Sci. USA. Vol. 95, pp. 4193-4198, April, 1998
23. S.A. Kauffman. The Origins of Order. Oxford University Press, Oxford, 1993
23a. M. Mitchell Waldrop, Complexity, Simon and Schuster, Touchstone, N.Y., 1992
24. Stan Bumble, Honig, J. M., Application of Order-Disorder Theory in Physical Adsorption, I. Fundamental Equations, J. Chem Physics, 33, 424, 1960
25. A. D. Sanchez, J. M. Lopez, and M. A. Rodriguez, Nonequilibrium Phase Transitions in Directed Small-World Networks, cond-mat/0110500.
26. M. E. J. Newman, S. H. Strogatz, D. J. Watts, Random Graphs with Arbitrary Degree Distributions and Their Applications, cond-mat/0007235
27. Kevin Bacon, The Small World, And Why It All Matters, Santa Fe Institute Bulletin, Vol. 14, Number 2, http://www.santafe.edu/sfi/publications/Bulletins/bulletinFall99/workInProgress/smallWorld.html
28. Watts, Duncan J., and Steven H. Strogatz. 1998. Collective dynamics of 'small-world' networks. Nature 393:440-442
29. Beom Jun Kim, Chang No Yoon, Seung Kee Han and Hawoong Jeong, Path Finding Strategies in Scale-Free Networks, cond-mat/0111232
30. J. Podani, Z. N. Oltvai, H. Jeong, B. Tombor, A.-I. Barabasi, and E. Szathmary, Comparable System-Level Organization of Archaea and Eukaryotes
31. Mads Ipsen and Alexander S. Mikhailov, Evolutionary Reconstruction of Graphs, nlin: AO/0111023
32. Ramon Ferrer i Cancho and Richard V. Sole, Optimization in Complex Networks, cond-mat/0111222
33. Lemonick, M. D., The Future of Drugs, The Labs: Inside the Brave new Pharmacy, Time Magazine, Special Issue, Jan. 15, 2001
34. Stan Bumble, Computer Generated Physical Properties, CRC Press, LLC, Boca Raton, 1999.
35. Stan Bumble, Computer Simulated Plant Design for Waste Minimization / Pollution Prevention, CRC Press LLC, Boca Raton, 2000
36. Friedler, F., Fan, L. T., Design of Molecules with Desired Properties by Combinatorial Analysis, 1997, Preprint
37. Ritter, E., THERM, User's Manual, Department of Chemical Engineering and Environmental Science, New Jersey Institute of Technology, Newark, 1980
38. Benson, S., Thermochemical Kinetics, John Wiley and Sons, N. Y., 1976
39. Engelman, D.A., Steitz, P., Goldman, A. 1986, Identifying nonpolar transbilayer helices in amino acid sequences of membrane proteins. Annu. Rev. Biophys. Chem. 15, 321-353
40. Zamyatnin, A. A., 1972. Protein volume in solution. Prog. Biophys. Mol. Biol. 24, 107-123
41. White, W. B., Johnson, S. M., and Dantzig: Chemical Equilibrium in Complex Mixtures, J. Chem. Phys., Vol. 28, 751-755, May, 1958
42. Fan, L. T., Friedler, F., Reaction Pathway Analysis by a Network Synthesis Technique, Annual Meeting, Session 264, AICHE

42a. Fan, L. T., B. Bertok, F. Friedler, A Graph-Theoretic Method to Identify Candidate Mechanisms for Deriving the Rate Law of a Catalytic Reaction, Computers and Chemistry 26 (2002) 265-292
42b. Seo, H., D.-Y. Lee, S. Park, L. T. Fan, S. Shafie, B. Bertok, F. Friedler, Biotechnology Letters 23: 1551-1557, 2001
43. S. Berkovich, The Meaning of DNA Information in the Phenomenon of Life, APS Virtual Pressroom-Centennial Meeting Lay Language Papers
44. Sheldrake, R., The Presences of the Past, Vintage, 1989
45. Sheldrake, R., The Rebirth of Nature, Parl Street Press, 1994